%% file: main.tex
\documentclass[conference]{IEEEtran}
\IEEEoverridecommandlockouts
\usepackage{cite}
\usepackage{amsmath,amssymb,amsfonts}
\usepackage{algorithmic}
\usepackage{graphicx}
\usepackage{textcomp}
\usepackage{xcolor}
\usepackage{amssymb}
\usepackage{times}
\usepackage{cases}
\usepackage{overpic}
\usepackage{caption}
\usepackage{subcaption}
\usepackage{multirow}
\usepackage{tabularx}
\usepackage{arydshln}
\usepackage{enumitem}
\usepackage{flushend}
\input{pkg_def}

\def\BibTeX{{\rm B\kern-.05em{\sc i\kern-.025em b}\kern-.08em
    T\kern-.1667em\lower.7ex\hbox{E}\kern-.125emX}}
\begin{document}

\title{A Deep Residual Star Generative Adversarial Network for multi-domain Image Super-Resolution}

\author{\IEEEauthorblockN{Rao Muhammad Umer}
\IEEEauthorblockA{\textit{Dept. of Computer Science} \\
\textit{University of Udine}\\
Udine, Italy \\
umer.raomuhammad@spes.uniud.it}
\and
\IEEEauthorblockN{Asad Munir}
\IEEEauthorblockA{\textit{Dept. of Computer Science} \\
\textit{University of Udine}\\
Udine, Italy \\
asad.munir@uniud.it}
\and 
\IEEEauthorblockN{Christian Micheloni}
\IEEEauthorblockA{\textit{Dept. of Computer Science} \\
\textit{University of Udine}\\
Udine, Italy \\
christian.micheloni@uniud.it}
}

\maketitle

\begin{abstract}
Recently, most of state-of-the-art single image super-resolution (SISR) methods have attained impressive performance by using deep convolutional neural networks (DCNNs). The existing SR methods have limited performance due to a fixed degradation settings, \ie usually a bicubic downscaling of low-resolution (LR) image. However, in real-world settings, the LR degradation process is unknown which can be bicubic LR, bilinear LR, nearest-neighbor LR, or real LR. Therefore, most SR methods are ineffective and inefficient in handling more than one degradation settings within a single network. To handle the multiple degradation, \ie refers to multi-domain image super-resolution, we propose a deep Super-Resolution Residual StarGAN (\srresstargan), a novel and scalable approach that super-resolves the LR images for the multiple LR domains using only a single model. The proposed scheme is trained in a StarGAN like network topology with a single generator and discriminator networks. We demonstrate the effectiveness of our proposed approach in quantitative and qualitative experiments compared to other state-of-the-art methods. 
\end{abstract}


\section{Introduction}
\label{sec:intro}
The goal of the single image super-resolution (SISR) is to reconstruct the high-resolution (HR) image from its low-resolution (LR) image counterpart. SISR problem is a fundamental low-level vision and image processing problem with various practical applications in satellite imaging, medical imaging, video enhancement and security and surveillance imaging as well. With the increasing amount of HR images / videos data on the internet, there is a great demand for storing, transferring, and sharing such large sized data with low cost of storage and bandwidth resources. Moreover, the HR images are usually downscaled to easily fit into display screens with different resolutions while retaining visually plausible information. The downscaled LR counterpart of the HR can efficiently utilize lower bandwidth, save storage, and easily fit to various digital displays. However, some details are lost and sometimes visible artifacts appear when users downscale and upscale the digital contents. 
\begin{figure}[t!]
\centering
\includegraphics[width=9.0cm]{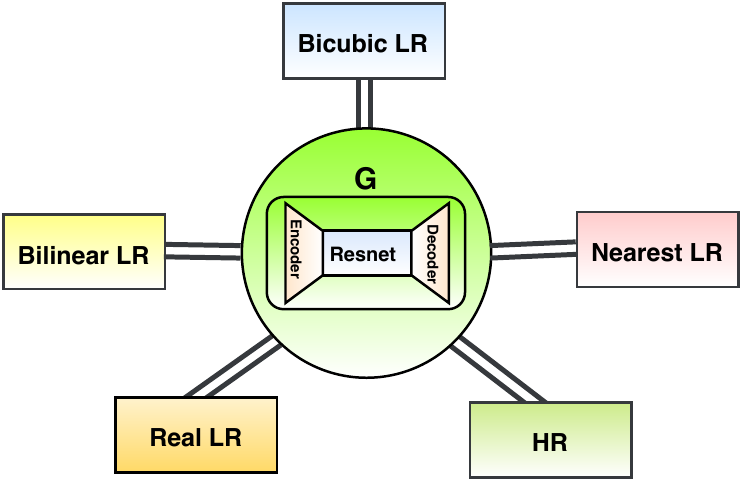}
\caption{The visualization of our proposed approach where a single generator $\mathbf{G}$ learns the mappings among multiple domains \ie LR/HR to LR/HR.}
\label{fig:star_tepology}
\vspace{-0.5cm}
\end{figure}

Usually, the SISR is described as a linear forward observation model~\cite{umer2020srrescgan,umer2020srrescycgan,Umer_2020_ICPR} by the following image degradation process:
\begin{equation}
    \by = \bH * \Tilde{\bx} + \eta,
    \label{eq:degradation_model}
\end{equation}
where $\by$ is an observed LR image, $\bH$ is a \emph{down-sampling operator} that convolves with an HR image $\Tilde{\bx}$ and resizes it by a scaling factor $s$, and $\eta$ is considered as an additive white Gaussian noise with standard deviation $\sigma$. However, in real-world settings,  $\eta$ also accounts for all possible errors during the image acquisition process that include inherent sensor noise, stochastic noise, compression artifacts, and the possible mismatch between the forward observation model and the camera device. In most of existing SR methods, the operator $\bH$ is usually known / fixed \ie bicubic. But in real-world settings, the operator $\bH$ is unknown that can be bicubic, bilinear, nearest, and real degradation kernel. 

Recently, numerous works have been addressed toward the task of SISR that are based on DCNNs for their powerful feature representation capabilities either on PSNR values~\cite{kim2016vdsrcvpr,Lim2017edsrcvprw,kai2017ircnncvpr,kai2018srmdcvpr,yuan2018unsupervised,Li2019srfbncvpr,zhang2019deep} or on visual quality~\cite{ledig2017srgan,wang2018esrgan,lugmayr2019unsupervised,fritsche2019dsgan,umer2020srrescgan,umer2020srrescycgan}. The SR methods mostly rely on the single degradation (\ie usually bicubic) with paired LR and HR images in the supervised training. In the real-world settings, the input LR image contains more complex degradation. As the bicubic, bilinear, and nearest LR degradation are rarely suitable for the real LR images, but these degradations can be used for data augmentation and are indeed a good choice for clean and sharp image super-resolution. Moreover, the \emph{Real LR} domain in the proposed method contains the LR/HR image pairs that follow the realistic physical image model instead of artificial ones. Due to such different degradation settings, the most SR methods often fail to produce convincing SR results or train their model independently for every pair of image domains. 

\begin{figure*}[h!]
\centering
\includegraphics[width=14cm]{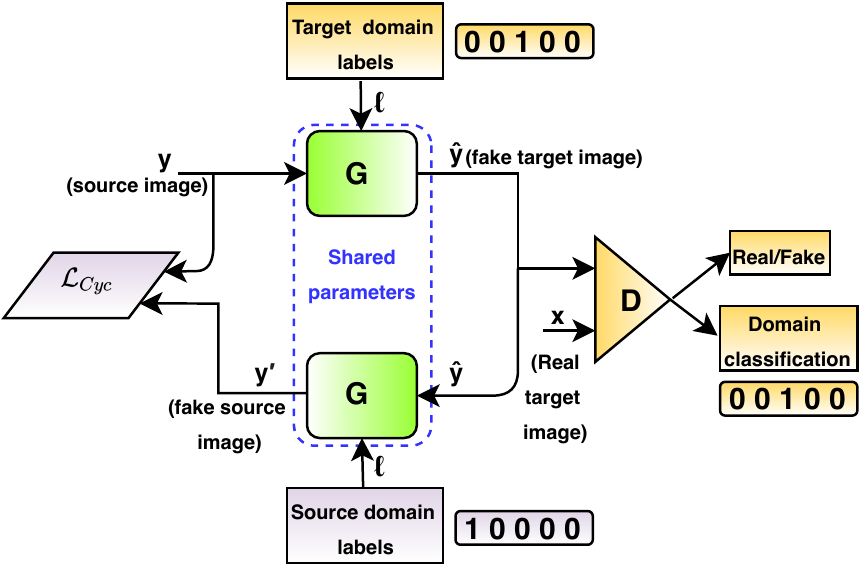}
\caption{Visualizes the structure of the our proposed \srresstargan~setup. $\mathbf{G}$ takes an input as both the source image ($\by$) and target domain
label and generates a fake target image ($\hat{\by}$). $\mathbf{G}$ tries to
reconstruct the fake source image ($\by^\prime$) from the fake target image ($\hat{\by}$) given the source domain label. $\mathbf{D}$ learns to discriminate between real and fake target images and classify the real target images to its corresponding domain. In this way, the $\mathbf{G}$ tries to generate fake target images indistinguishable from real target images and classifiable as target domain images by the $\mathbf{D}$.}
\label{fig:srrescstargan_arch}
\vspace{-0.5cm}
\end{figure*}
Instead of learning a fixed degradation setting (e.g., bicubic LR), the proposed \srresstargan~ takes as inputs both an  image and a domain label, and
learns to generate the image into the corresponding domain (LR/HR). The domain labels are encoded in binary or one-hot vector to represent domain information. Fig.~\ref{fig:star_tepology} illustrates an overview of our proposed SISR approach, where a single network $\mathbf{G}$ learns the mappings among the multiple domains \ie Bicubic LR, Bilinear LR, Nearest LR, Real LR, and HR. During the training phase, we translate the source domain images into the target domain images by randomly generate target domain labels. At the testing phase, we fix the target domain as HR domain to generate the SR images from any blind LR domain. The proposed scheme is inspired by the recent success of StarGAN~\cite{choi2018stargan} for multi-domain image-to-image translation applications. The proposed approach (refers to the section~\ref{sec:proposed_method} for more details) allows us to simultaneously train for multiple domains with in a unified network.  

Our contributions in this paper are as follows: 
\begin{itemize} 
 \item  We propose \srresstargan~ for multi-domain image super-resolution task. In contrast to the existing deep SISR methods, our method learns the mappings among multiple domains with in a single model. 
 \item The proposed scheme requires a single generator and discriminator networks to train efficiently for the images of multiple domains. 
 \item We provide both qualitative and quantitative results on SR benchmarks to show its superiority over existing SR methods.
\end{itemize}

\section{Related Work}
Recently, numerous works addressed the task of SISR using deep CNNs for their powerful feature representation capabilities. A preliminary CNN-based method to solve SISR is a super-resolution convolutional network with three layers (SRCNN)~\cite{dong2014srcnneccv}. Kim~\etal\cite{kim2016vdsrcvpr} proposed a very deep SR (VDSR) network with a residual learning approach. Lim \etal\cite{Lim2017edsrcvprw} proposed an enhanced deep SR (EDSR) network by taking advantage of the residual learning. In \cite{srwdnet}, the authors proposed SRWDNet to solve the joint deblurring and super-resolution task by following the realistic degradation. In \cite{Umer_2020_ICPR}, the authors proposed the ISRResCNet to solve the SISR in an iterative manner. These methods mostly rely on the PSNR-based metric by optimizing the $\mathcal{L}_1$/$\mathcal{L}_2$ losses in a supervised way by the LR/HR paired data.

For the perception SR task, a preliminary attempt was made by Ledig \etal\cite{ledig2017srgan} who proposed the SRGAN method to produce perceptually more pleasant results. To further enhance the performance of the SRGAN, Wang \etal\cite{wang2018esrgan} proposed the ESRGAN model to achieve the state-of-art perceptual performance. Despite their success, the previously mentioned methods are trained with LR/HR image pairs with the bicubic downsampling and thus they have limited performance in the real-world settings. Recently, in the real-world SR challenge series~\cite{AIM2019RWSRchallenge,NTIRE2020RWSRchallenge,AIM2020_RSRchallenge}, the authors have described the effects of bicubic downsampling. More recently, in~\cite{umer2020srrescgan,umer2020srrescycgan}, the authors proposed  GAN-based SR methods to solve the real-world SR problem. However, the above SR methods have different models to be independently built for every LR degradation. Our approach takes into account the different LR degradation within a single network by greatly increasing its applicability in practical scenarios. 

\section{Proposed Method}
\label{sec:proposed_method}
\subsection{Multi-domain training strategy}
Our goal is to train a single network $\mathbf{G}$ that super-resolves the LR images from multiple LR domains. To achieve this objective, we train the generator $\mathbf{G}$ to translate an input source image $\by$  into an output target image $\hat{\by}$ conditioned on the target domain labels $\mathbf{\ell}$ as $\mathbf{G}(\by, \mathbf{\ell}) \rightarrow \hat{\by}$. The domain labels $\mathbf{\ell}$ are encoded in binary or one-hot vector to represent domain information. We randomly generate the target domain labels $\mathbf{\ell}$ so that the network $\mathbf{G}$ flexibly translates the source domain images into the target domain images. The discriminator network $\mathbf{D}$ learns to distinguish between real/fake images and also tries to minimize the domain classification error only associated with the known domain label. The discriminator $\mathbf{D}$ (\ie adapted from StarGAN~\cite{choi2018stargan}) produces the probability distributions over both target images and target domain labels, $\mathbf{D}: \bx \rightarrow \{\mathbf{D}_{trg}(\bx), \mathbf{D}_{cls}(\bx)\}$. Fig.~\ref{fig:srrescstargan_arch} illustrates the training strategy of the proposed \srresstargan. 
 
\subsection{Network Architectures}
\label{sec:net_arch}
\textbf{Generator ($\mathbf{G}$):}
We use the generator $\mathbf{G}$ network as a \emph{Encoder-Resnet-Decoder} like structure, adapted from SRResCGAN~\cite{umer2020srrescgan} which strictly follows the degradation process~\eqref{eq:degradation_model}. In the $\mathbf{G}$ network, both \emph{Encoder} and \emph{Decoder} layers have $64$ convolutional feature maps of $5\times5$ kernel size with $C \times H\times W$ tensors, where $C$ is the number of channels of the input image. \emph{Resnet} consists of $5$ residual blocks with two Pre-activation \emph{Conv} layers, each of $64$ feature maps with kernel support $3\times3$, and the preactivation is the \emph{Sine} nonlinearities layer with $64$ output feature channels. The trainable projection layer~\cite{Lefkimmiatis2018UDNet} inside the \emph{Decoder} computes the proximal map with the estimated noise standard deviation $\sigma$ and handles the data fidelity and prior terms.\\
\textbf{Discriminator ($\mathbf{D}$):}
The discriminator network ($\mathbf{D}$) learns to distinguish between real and fake images and classify the real images to its corresponding domain. The  discriminator network contains 6 convolutional layers with kernels that support $4\times4$ of increasing feature maps from $64$ to $2048$ followed by the leaky ReLU as done in StarGAN~\cite{choi2018stargan}.

\subsection{Network Losses}
\label{sec:net_losses}
To train the generator $\mathbf{G}$ and discriminator $\mathbf{D}$, we use the following respective objective loss functions:
\begin{equation}
    \mathcal{L}_{G} = \mathcal{L}_{\mathrm{per}}+ \mathcal{L}_{\mathrm{GAN}} + \mathcal{L}_{tv} + \mathcal{L}_{cls}^{f} + 10\cdot \mathcal{L}_{\mathrm{1}} + 10\cdot \mathcal{L}_{\mathrm{cyc}},
    \label{eq:l_g}
\end{equation}
\begin{equation}
\mathcal{L}_{D} =  \mathcal {L}_{GAN} +  \mathcal{L}_{cls}^{r}, 
\label{eq:l_d}
\end{equation}
where $\mathcal{L}_{per}$, $\mathcal{L}_{GAN}$, $\mathcal{L}_{tv}$, $\mathcal{L}_{cls}^{r/f}$, $\mathcal{L}_{1}$, and $\mathcal{L}_{cyc}$ denote the perceptual (VGG-based) loss, texture/GAN loss, total variation loss, real/fake classification loss, content loss, and cyclic loss respectively.

\section{Experiments}
\subsection{Training data}
\label{sec:train_data}
For the training, we use DIV2K~\cite{div2k}, Flickr2K~\cite{timofte2017flickr}, and RealSR~\cite{wei2020aim_realSR} datasets that jointly contain 22,430 high-quality HR images for image restoration tasks with rich and diverse textures. We obtain the LR bicubic, LR bilinear, and LR nearest images by down-sampling HR images by the scaling factor $\times4$ of the DIV2K and Flickr2K datasets using the Pytorch bicubic, bilinear, and nearest kernel function. The LR real images by the scaling factor $\times4$ are provided with their corresponding HR images in the RealSR dataset. We consider the five domains as shown in Fig.~\ref{fig:star_tepology}. We obtain the source domain images from the three datasets with their corresponding domain labels. We randomly generate the target domain labels with their corresponding images from the three datasets. We train our network in RGB color space. 

\subsection{Data augmentation}
We augment the training data with random vertical and horizontal flipping, and $90^{\circ}$ rotations. Moreover, we also consider another effective data augmentation technique, called  mixture of augmentation (MOA) \cite{yoo2020rethinking} strategy. In the MOA, a data augmentation (DA) method, among \ie Blend, RGB permutation, Mixup, Cutout, Cutmix, Cutmixup, and CutBlur is randomly selected and then applied on the inputs. This MOA technique encourages the network to acquire more generalization power by partially blocking or corrupting the training sample. 

\subsection{Training details}
\label{sec:net_training}
At the training time, we set the input image patch size as $128\times128$. We train the network for 51000 training iterations with a batch size of 16 using Adam optimizer with parameters $\beta_1 =0.9$, $\beta_2=0.999$, and $\epsilon=10^{-8}$ without weight decay for both generator and discriminator to minimize the loss functions in the Eqs.~\eqref{eq:l_g} and \eqref{eq:l_d}. The learning rate is initially set to $10^{-4}$ and then multiplies by $0.5$ after 5K, 10K, 20K, and 30K iterations. Our all experiments are performed with a scaling factor of $\times4$ for the LR and HR images.
\begin{table*}[h!]
	\centering
	\caption{$\times4$ SR quantitative results comparison of our method with others over the DIV2K (100 images of validation-set) and RealSR (93 images of testset) that are total 393 images of the testset with the four LR degradation. The arrows indicate if high $\uparrow$ or low $\downarrow$ values are desired. The best performance is shown in {\color{red} red} and the second best performance is shown in {\color{blue} blue}.}
	\tabcolsep=0.01\linewidth
	\scriptsize
	\resizebox{1.0\textwidth}{!}{
	\begin{tabular}{|c|c|c|c|c|c|c|}
	    \hline
	    \multirow{2}{*}{\textbf{SR methods}} & \multirow{2}{*}{\textbf{\#Params}} & 
	    \multicolumn{1}{c|}{\textbf{Bicubic}} & \multicolumn{1}{c}{\textbf{Bilinear}} &
	    \multicolumn{1}{|c|}{\textbf{Nearest}} &
	    \multicolumn{1}{c|}{\textbf{Real}} &
	    \multicolumn{1}{c|}{\textbf{Average}} \\
	    \cline{3-7}
		 &  & PSNR$\uparrow$ / SSIM$\uparrow$ / LPIPS$\downarrow$ & PSNR$\uparrow$ / SSIM$\uparrow$ / LPIPS$\downarrow$ & PSNR$\uparrow$ / SSIM$\uparrow$ / LPIPS$\downarrow$ & PSNR$\uparrow$ / SSIM$\uparrow$ / LPIPS$\downarrow$ & PSNR$\uparrow$ / SSIM$\uparrow$ / LPIPS$\downarrow$ \\ 
		\hline
		EDSR~\cite{Lim2017edsrcvprw} & $43M$ & 21.33 / 0.66 / 0.3477 & 23.05 / 	0.72 / {\color{blue}0.3083} & 19.34 / 0.56 / 0.3653 & {\color{red}28.06} / {\color{red}0.82} / 0.4182  & 22.95 / 0.69 / 0.3599  \\
		ESRGAN~\cite{wang2018esrgan} & $16.7M$ & 16.02 / 0.31 / 0.6008 & 17.25 / 0.37 / 0.5186 & 15.29 / 0.26 / 0.6254 & 27.98 / {\color{red}0.82} / 0.3840  & 19.14 / 0.44 / 0.5322 \\
		SRResCGAN~\cite{umer2020srrescgan} & $380K$ & 23.30 / 0.67 / {\color{red}0.2900} & 24.43 / 0.70 / {\color{red}0.2720} & 21.10 / 0.60 / {\color{red}0.3044} & 27.96 / {\color{red}0.82} / {\color{red}0.3676} & 24.20 / 0.69 / {\color{red}0.3085} \\
		SRResCycGAN~\cite{umer2020srrescycgan} & $380K$ & {\color{blue}24.56} / {\color{blue}0.73} / {\color{blue}0.3380} & {\color{blue}25.58} / {\color{blue}0.75} / 0.3183 & {\color{blue}21.75} / {\color{blue}0.63} / {\color{blue}0.3541} & {\color{blue}28.02} / {\color{red}0.82} / {\color{blue}0.3827}  & {\color{blue}24.98} / {\color{blue}0.73} / {\color{blue}0.3483} \\
		\srresstargan~ (ours) & $380K$ & {\color{red}25.52} / {\color{red}0.75} / 0.4250 & {\color{red}26.23} / {\color{red}0.76} / 0.4083 & {\color{red}22.75} / {\color{red}0.66} / 0.4415 & {\color{blue}28.02} / {\color{red}0.82} / 0.4353  & {\color{red}25.63} / {\color{red}0.75} /	0.4275\\
		\hline
	\end{tabular}}
	\label{table:comp_sota}
	\vspace{-0.2cm}
\end{table*}
\subsection{Evaluation metrics}
We evaluate the trained model under the Peak Signal-to-Noise Ratio (PSNR), Structural Similarity (SSIM), and LPIPS~\cite{zhang2018unreasonable} metrics. The PSNR and SSIM are distortion-based measures, while the LPIPS better correlates with human perception. We measure the LPIPS based on the features of pretrained AlexNet~\cite{zhang2018unreasonable}. The quantitative SR results are evaluated on the $RGB$ color space.         
\begin{figure*}[htbp!]
    \centering
    \begin{subfigure}[t]{1.0\textwidth}
        \centering
        \includegraphics[width=14.0cm]{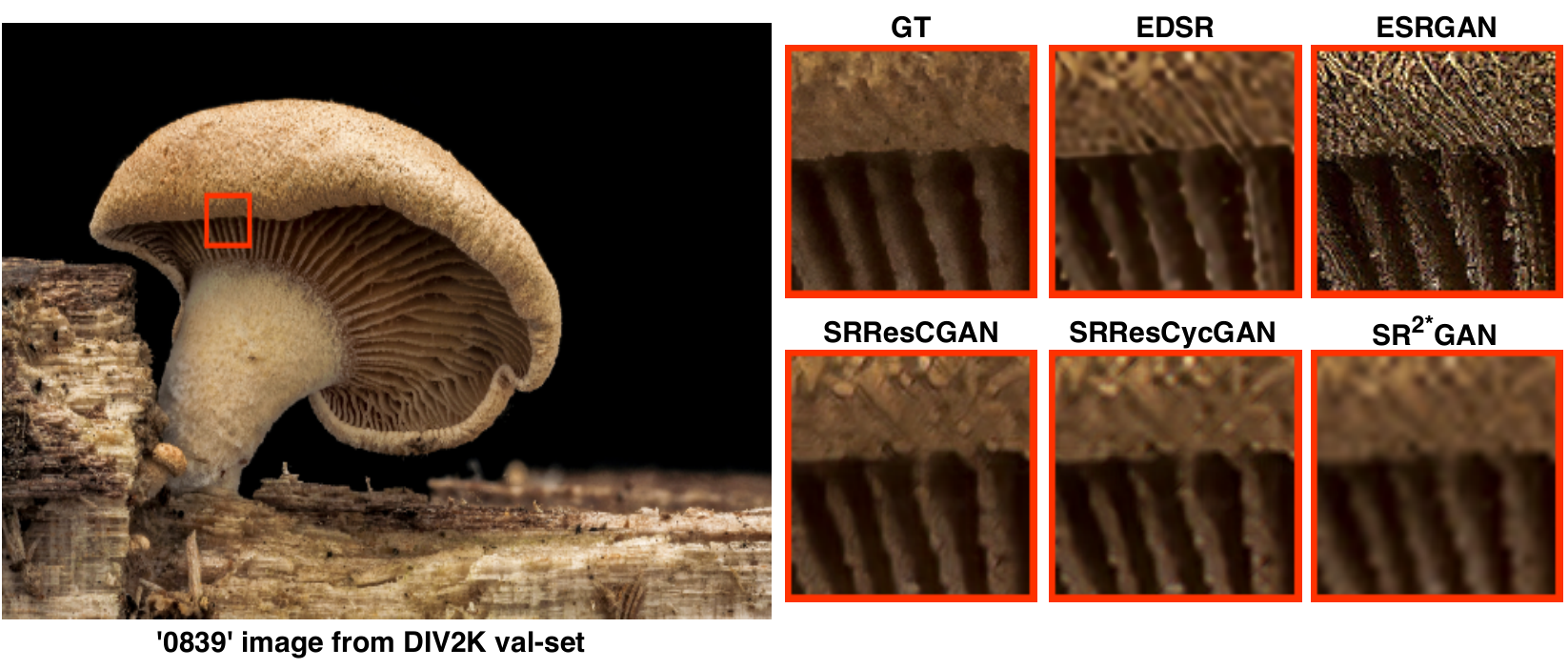}
        \caption{SR results on the LR bicubic}
    \end{subfigure}\\ 
    \begin{subfigure}[t]{1.0\textwidth}
        \centering
        \includegraphics[width=14.0cm]{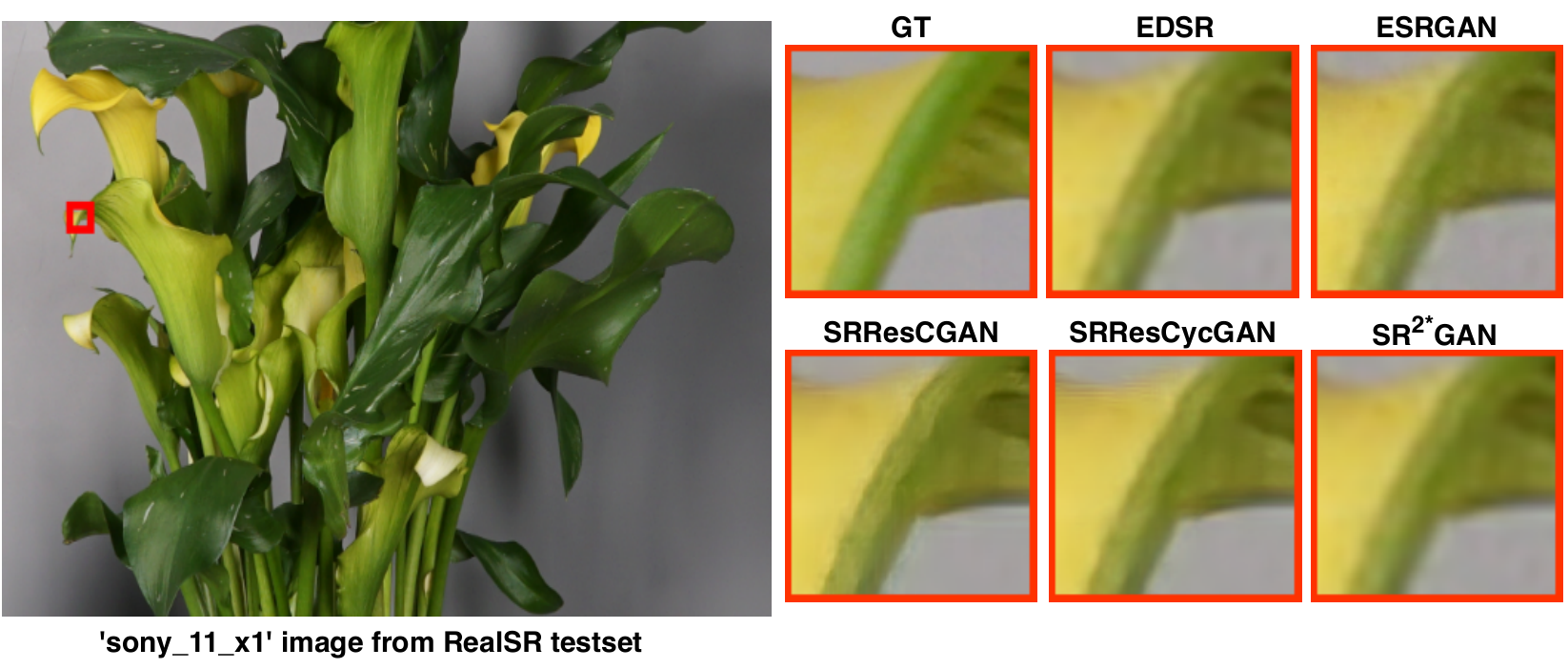}
        \caption{SR results on the LR real}
    \end{subfigure}
    \caption{Visual comparison of our method with other state-of-art methods on $\times4$ super-resolution.}
    \label{fig:4x_result}
    \vspace{-0.5cm}
\end{figure*}

\subsection{Comparison with the state-of-art methods}
\label{sec:comparison with sota}
We compare our method with other state-of-art SISR methods including EDSR~\cite{Lim2017edsrcvprw}, ESRGAN~\cite{wang2018esrgan}, SRResCGAN~\cite{umer2020srrescgan}, and SRResCycGAN~\cite{umer2020srrescycgan}. We compare the SR results with one deep feed-forward residual network (\ie EDSR) and other three GAN-based approaches (\ie ESRGAN, SRResCGAN, SRResCycGAN). Although there are existing SR methods \cite{kai2018srmdcvpr}, \cite{zhang2019deep}, and \cite{srwdnet} by handling multiple degradation settings, but they are the non-blind SR techniques with deep feed-forward networks, while our proposed method is a blind SR GAN-based approach. So, the comparison with \cite{kai2018srmdcvpr}, \cite{zhang2019deep}, and \cite{srwdnet} methods is not fair. We run all the original source codes and trained models by the default parameters settings for the comparison. The competing methods are trained for just one degradation setting and tested on the different ones. Unlike them, the proposed approach can be trained with multiple degradations instead of single degradation.

Table~\ref{table:comp_sota} shows the quantitative results comparison of our method with the others over the DIV2K (100 images of validation-set) and RealSR (93 images of testset) that are total 393 images of the testset with the four LR degradation. We have excellent results in terms of PSNR and SSIM compared to the other methods, while in the case of LPIPS, we lag by the others because our training objective is to get better PSNR/SSIM. The reason for none of the competing methods achieved their best performance against the bicubic down-sampling, is that the other GAN-based approaches focus more on the perceptual quality, as good perceptual image quality reduces the PSNR/SSIM scores. Our model is trained with reasonably perceptual quality, while good PSNR/SSIM, refers to section~\ref{sec:abl_study}. 

Regarding the visual quality, Fig.~\ref{fig:4x_result} shows the visual comparison of our method with other SR methods at the $\times 4$ upscaling factor on the test-set. We have comparable visual SR results with others. 

\begin{table}[!ht]
	\centering
	\caption{This table reports the quantitative SR results of our method over the DIV2K and RealSR validation-set (20 images, not used during the training phase) with the four LR domains (\ie Bicubic, Bilinear, Nearest, Real) for our ablation study. The arrows indicate if high $\uparrow$ or low $\downarrow$ values are desired.}
	\tabcolsep=0.01\linewidth
	\scriptsize
	\resizebox{0.50\textwidth}{!}{
	\begin{tabular}{|l|c|c|c|c|}
		 \hline
		 \textbf{SR method} & \textbf{Domain label Conditioning ($\ell$)} & \textbf{PSNR$\uparrow$} & \textbf{SSIM$\uparrow$} & \textbf{LPIPS$\downarrow$} \\ 
		\hline
		\srresstargan~ (v1) & {\begin{tabular}[c]{@{}c@{}}~$\mathbf{G}$ Conditioned on only HR target domain~ \\ \& used two separate Disc.~($\mathbf{D}_{trg}$, $\mathbf{D}_{src}$)\end{tabular}} & 26.95 & 0.75 & 0.3423  \\
		\hline
		\srresstargan~ (v2) & {\begin{tabular}[c]{@{}c@{}}~$\mathbf{G}$ Conditioned on only HR target domain~ \\ \& used one Disc.~($\mathbf{D}_{trg}$)\end{tabular}}  & 27.20 & 0.76 & \textbf{0.3176}  \\
		\hline
		\srresstargan~ (v3) & {\begin{tabular}[c]{@{}c@{}}~$\mathbf{G}$ Conditioned on random target domains~ \\ \& used one Disc.~($\mathbf{D}_{trg}$)\end{tabular}}  & \textbf{28.09} & \textbf{0.78} & 0.3891  \\
		\hline
	\end{tabular}}
	\vspace{-0.5cm}
	\label{table:ablation_study}
\end{table}
\subsection{Ablation Study}
\label{sec:abl_study}
For our ablation study, we train three variants of the proposed \srresstargan~ network structure with conditioning of the domain labels. By referencing to the Fig.~\ref{fig:srrescstargan_arch}, the first network variant v1 is trained by conditioning the first branch generator $\mathbf{G}$ on only HR target domain labels, while, the other branch generator  $\mathbf{G}$ (shared weights) is conditioned on the multiple LR source domain labels. We use two separate discriminators \ie $\mathbf{D}_{trg}$ for the target domain supervision, $\mathbf{D}_{src}$ for the source domain supervision in the \srresstargan~ (v1). Similarly, the second network variant v2 is trained as the v1, but only uses one discriminator \ie $\mathbf{D}_{trg}$ for the target domain supervision. Finally, the third network variant v3 is trained by conditioning both generators $\mathbf{G}$ on random domain labels and use only one discriminator $\mathbf{D}_{trg}$ for the real target domain supervision by minimizing the the total loss functions in the Eqs.~\eqref{eq:l_g} and \eqref{eq:l_d}. Table~\ref{table:ablation_study} shows the quantitative SR results of \srresstargan~ (v1, v2, v3) over the DIV2K and RealSR validation-sets (20 images, not used during the training phase) with the four LR domains \ie Bicubic, Bilinear, Nearest, Real. The third network variant v3 performs better in terms of PSNR/SSIM among others. This shows the effectiveness of random target domain labeling and real target domain images supervision by the discriminator. Therefore, we opt for the third version v3 and used it for the evaluation in the section-\ref{sec:comparison with sota}.

\section{Conclusion}
We proposed a unified deep \srresstargan~ for the multi-domain image super-resolution task. The proposed scheme learns the multiple degradation of different domains within a single model. The proposed method trains the single generator and discriminator networks to efficiently learn mappings among multiple domains using StarGAN like network topology. Our method achieves excellent SR results in terms of the PSNR/SSIM values compared to the existing SR methods.  

\section*{Acknowledgement}
This work was supported by EU H2020 MSCA through Project ACHIEVE-ITN (Grant No 765866).

\bibliographystyle{IEEEtran}
\bibliography{IEEEfull,refs}

\end{document}

%% file: pkg_def.tex
\def\bx{{\bf x}}
\def\by{{\bf y}}

\def\0{{\bf 0}}
\def\1{{\bf 1}}

\def\bH{{\bf H}}

\def\etal{\emph{et al. }}
\def\ie{\emph{i.e. }}

\def\srresstargan{SR$^{2~*}$GAN}